\documentclass[12pt,a4paper]{article}
\usepackage{amssymb,amsfonts,amsmath}
\usepackage{epsfig, euscript, graphics}

\begin{document}

\title{Bohm-Bell type experiments:  Classical probability approach to (no-)signaling and applications to quantum physics and psychology}
\author{Andrei Khrennikov$^{1,2}$,   Alexander Alodjants$^{1}$}

\maketitle

$^1$ National Research University for Information Technology, Mechanics and Optics (ITMO), department, St.
Petersburg, 197101, Russia

$^2$  Linnaeus University, International Center for Mathematical Modeling in Physics and Cognitive Sciences, V\"axj\"o,
SE-351 95, Sweden

\begin{abstract}We consider the problem of representation of quantum states and observables in the framework of classical probability theory 
(Kolmogorov's measure-theoretic axiomatics, 1933). Our aim is to show that, in spite of  the common opinion, correlations of observables 
$A_1, A_2$ and $B_1,B_2$ involved in the experiments of the Bohm-Bell type can be expressed as correlations 
of classical random variables $a_1, a_2$ and $b_1, b_2.$
The crucial point is that correlations $\langle A_i, B_j \rangle$ should be treated as conditional on the selection of the pairs  $(i, j).$  
The setting selection procedure is based on two random generators $R_A$ and $R_B.$ They are also considered as observables, 
supplementary to the ``basic observables''    $A_1, A_2$ and $B_1, B_2.$ These observables are absent in the standard 
description, e.g., in the scheme for derivation of the CHSH-inequality. We represent them by classical random variables 
$r_a$ and $r_b.$  Following the recent works 
 of Dzhafarov and collaborators, we apply our conditional correlation approach to characterize  (no-)signaling in the classical 
probabilistic framework.  Consideration the Bohm-Bell experimental scheme in the presence of signaling is important for applications outside quantum mechanics, e.g., in psychology
and social science.  
\end{abstract}

{\bf keywords}: quantum versus classical probability; Bohm-Bell type experiments in physics and psychology; (no-)signaling; random generators; conditional probability.

\section{Introduction}

This paper is related to the old foundational problem of quantum mechanics: {\it whether it is possible to represent quantum states by classical probability (CP) distributions and 
quantum observables by random variables.} (In fact, we analyze the general measurement scheme involving compatible and 
 incompatible observables which need not be described by the quantum formalism. But our starting point is construction of the CP-representation for 
quantum mechanics.) 

\subsection{Towards CP-representation}

The first step to study the problem of CP-embedding of quantum mechanics was done by Wigner \cite{W} who tried to construct the joint probability distribution (jpd) of the position and 
momentum observables. However, Wigner's function can take negative values. We also mention the Husimi-Kano Qfunction  \cite{H, K}  and Glauber-Sudarshan 
function  \cite{G, S}. These functions are widely applied to quantum mechanics and field theory (e.g., \cite{Brida}). However, they cannot be described by CP-theory. 
The first CP-representation  of quantum mechanics based on of {\it symplectic tomogram}  was constructed in  \cite{MA1} (see also \cite{MA2, MA3}).  

 Another construction of the CP-representation of quantum mechanics is  based on so-called {\it prequantum classical statistical field theory} 
\cite{KHRP0a}-\cite{KHRP2}. In this theory 
quantum states (density operators) are represented by covariation operators of random fields valued in complex Hilbert space $H$ (the state space of the quantum formalism); quantum 
observables (Hermitian operators) are represented by quadratic forms of such fields.  The  classical$\to$quantum map has the form:
\begin{equation}
\label{PC} 
B \to \rho= B/\rm{Tr} B, \; f_A \to A, \mbox{for the quadratic form} \; f_A( \phi)=\langle \phi \vert A\vert \phi \rangle.
\end{equation} 
This  representation suffers of violation of the spectral postulate: the range of values of a quadratic form differs from the spectrum 
of the corresponding operator.  To solve this problem, prequantum classical statistical field theory was completed by the corresponding measurement theory based on the detectors 
of the threshold type \cite{KHRP3a, KHRP3b}. However, the majority of physicists would not be convinced that quantum effects can be reduced to behavior
 of classical random field combined with 
threshold detection (although experimenters typically recognize the role of detection thresholds in quantum measurements). 

It should be honestly said  the tomographic and random field approaches 
were practically ignored by the quantum foundational community.   Nowadays it is commonly 
believed that CP-theory, see Kolmogorov \cite{K}, cannot serve to represent quantum observables. The roots of this belief lie in the previous 
unsuccessful attempts to construct the CP-representation for quantum mechanics, starting with Wigner's attempt. 

\subsection{Bell's type no-go statements}

However, the main argument against the possibility to proceed  with the CP-representation is based on {\it no-go theorems.} The first no-go theorem was proven by
 von Neumann \cite{VN} (German edition -1933): the theorem on nonexistence of dispersion free 
states. This theorem was strongly criticized   by Bell \cite{BE2} who pointed to non-physicality of von Neumann's rule for correspondence between classical and quantum probabilistic 
structures (probabilities$\to$states, random variables$\to$ Hermitian operators), cf.  sections \ref{CC1}, \ref{CC2}.  Bell's own no-go theorem \cite{BE1, BE2}has much better reputation 
than von Neumann's theorem and it has the very big impact to quantum foundations, quantum information, and quantum technology  
(at the same time it generated a plenty of critical papers, see, e.g.,  \cite{BC1}-\cite{BC4} for some resent publications).  Bell proposed the CP-description of the Bohm-Bell type experiments.  
This approach is known as the hidden variables description. Since it is very difficult  to test experimentally the original Bell inequality (see \cite{L8, L9} for a discussion),
Clauser, Horne, Shimony, \& Holt (CHSH) \cite{CHSH} modified Bell's approach on the basis of the CHSH-inequality. (In spite of a rather common opinion,
this modification does not equivalent to the original Bell approach.) We denote the CP-model proposed by them 
by the symbol ${\cal M}_{BCHSH}$ (see section \ref{RV}).  

Bell emphasized the role of {\it nonlocality} \cite{BE1, BE2}. However, Fine \cite{Fine1, Fine2} showed that
the    CHSH-inequality is satisfied if and only if the assumption on  the existence of the jpd for the four observables $A_1, A_2, B_1, B_2$ involved in the experiment, see section \ref{EPL}. 
The latter is equivalent to using CP-theory. Therefore a violation of  the    CHSH-inequality inequality by quantum (theoretical and experimental) probabilities 
implies inapplicability of CP-theory.  Erroneously inapplicability of one concrete CP-model, namely, ${\cal M}_{BCHSH},$ to 
describe the Bohm-Bell type experiments was commonly treated as inapplicability of CP in general.  

Nevertheless,  as was shown by Khrennikov and coauthors \cite{KHRB1},  \cite{KHRB2} and  recently by Dzhafarov and coauthors \cite{D1}-\cite{D7},  
the Bohm-Bell type experiments can be modeled with the aid of the CP-representation of quantum  observables. However, such  
CP-models are not so straightforward as  ${\cal M}_{BCHSH}.$ Denote the models developed in  \cite{KHRB2} and in  \cite{D1}-\cite{D7} by the symbols
${\cal M}_{KH}$ and ${\cal M}_{DZ},$ respectively.    

\subsection{Conditional probability approach}

The basic distinguishing feature of ${\cal M}_{KH}$  is  taking into account the {\it conditional nature} of quantum probabilities. Generally,  we follow Ballentine \cite{BL, BL1}, 
especially his paper \cite{BL2}.  In the previous works  \cite{KHRB1},  \cite{KHRB2} and the present paper conditioning is modeled with the aid of 
the {\it  random generators  selecting the experimental settings.} They are  represented as 
random variables (RVs) $r_a, r_b$ which are supplementary to the ``basic'' RVs $a_1, a_2, b_1, b_2$ (see sections \ref{RV}, \ref{RV1}).  
These  RVs are absent in  ${\cal M}_{BCHSH}.$ At the same time the random generators play the crucial role in the real experimental design 
of such experiments. We remark that Bohr emphasized that in modeling quantum phenomena all components of the experimental   arrangement should 
be taken into account \cite{BR, PL}. Thus ignoring the random generators makes a model without them (as, e.g., ${\cal M}_{BCHSH})$  inadequate to the real physical situation.  

However, model   ${\cal M}_{BCHSH}$ need not be rigidly coupled to  conditioning on selection of experimental settings. A variety of conditioning can lead to violation of 
the Bell type inequalities. In particular, in the random field theory \cite{KHRP3a, KHRP3b} endowed with the threshold detection scheme this is conditioning on joint detection or more generally 
on a time widow for coupling two clicks of spatially separated detectors, see \cite{OSID} for the general discussion.  

We remark that CP-conditioning is one of the forms  of the mathematical representation of context-dependence, dependence of outputs of observables on components of 
the experimental context (see again Bohr \cite{BR}). Thus ${\cal M}_{KH}$ can be treated as a contextual model of the Bohm-Bell type experiments.  However, one has to be 
very careful with the use of the notion ``contextuality''. Here it matches the Copenhagen interpretation of quantum mechanics in its original Bohr's understanding 
\cite{PL}, cf. with the notion of  ``quantum contextuality'' based on the Bell type tests. 
In contrast to the latter, Bohr's type contextuality is not mystical - it is straightforwardly  coupled to the experimental 
arrangement.  We can also move another way around: to start with a general contextual probabilistic model and then to describe the class of such models which can be represented in complex Hilbert 
space $H,$ see \cite{KHRC1}-\cite{KHRC} ({\it ``constructive wave function approach''}).

\subsection{CP-representations in the presence of signaling}

Model ${\cal M}_{DZ}$ does not contain explicit counterparts of the random generators for setting's selection. It is based on contextual  coupling of random variables corresponding 
to the choice of experimental settings. In spite of different mathematical structures, both models,  ${\cal M}_{KH}$ and ${\cal M}_{DZ},$ represent the procedure of 
experimental settings' selection:  ${\cal M}_{KH}$  with the aid of the random generators,  ${\cal M}_{DZ}$  with the aid of  contextual indexing of RVs representing observables.

Model  ${\cal M}_{DZ}$ was applied to study {\it  contextuality} in the CP-framework with the especial emphasis of the possibility to proceed 
in the presence of {\it signaling} \cite{D1}-\cite{D7}. (Contextuality studied by Dzhafarov and the coauthors is the natural
 extension of the notion of quantum  contextuality based on the Bell type tests.)   We remark that signaling is absent in quantum mechanics. 
Therefore contextuality theory developed in \cite{D1}-\cite{D7} and known as contextuality by default (CbD) is more general than the  standard 
theory of quantum contextuality. In particular, the standard Bell type inequalities are modified  by including the signaling contribution.
They are known as the {\it Bell-Dzhafarov-Kujala} (BDK) inequalities.   This generality provides the possibility to apply CbD outside  physics, especially in 
psychology \cite{BPS1}-\cite{EDSQ}, 
where the condition of no-signaling is generally violated \cite{D3, D6}.   

Papers \cite{KHRB1},  \cite{KHRB2} were aimed to show the existence of the CP-representation for 
the Bohm-Bell experiment with genuine quantum systems. In these papers  model 
${\cal M}_{KH}$ was presented in the very concrete framework coupled to classical versus quantum discussion on the CHSH-inequality.
This rigid coupling with quantum mechanics led to ignoring the possibility to use model ${\cal M}_{KH}$ even in the presence of signaling.
Consistent CP-treatment of (no-)signaling  in model ${\cal M}_{DZ}$ motivated the authors of this paper to analyze (no-)signaling issue on the basis
of ${\cal M}_{KH}.$ And we found very clear CP-interpretation of no-signaling: {\it independence of  RVs $a_1, a_2, r_a$ representing Alice's observables and random generator 
from  RV $r_b$ representing the random generator for  selecting Bob's observables. }Thus no-signaling has clear probabilistic  meaning. 

In contrast to papers \cite{KHRB1},  \cite{KHRB2}, in this paper we proceed in very general abstract framework which can be used both in physics and outside it, e.g., in psychology.
(See \cite{BPS1}-\cite{EDSQ} for consideration of the Bell type  inequalities in psychology.)

Finally, we point to recently published paper of Margareta Manjko and Vladimir Manjko  \cite{MA4} 
presenting a very general scheme of the CP-representation of quantum states and observables.

\section{Bohm-Bell  type experiment: traditional description}

\subsection{Description of (four) observables}
\label{EPL}

In the observational framework for the Bohm-Bell  type experiments,  there are considered {\it four observables} $A_1, A_2, B_1, B_2$ taking values $\pm 1.$ 
It is assumed that the pairs of  observables $(A_i, B_j), i, j=1,2,$ 
can be measured jointly, i.e., $A$-observables are compatible with $B$-observables. 
However, the observables in pairs $A_1, A_2$ and $B_1,B_2$ are incompatible, i.e., they cannot be jointly measured. Thus probability 
distributions $p_{A_i B_j}$ are well defined theoretically by quantum mechanics and they can be verified experimentally; probability distributions $p_{A_1 A_2}$ and $p_{B_1 B_2}$
are not defined by quantum mechanics and, hence, the question of their experimental verification does not arise.  

We stress that, although our starting point is quantum mechanics and the Bohm-Bell experiment for measurement of spin of electrons or polarization of photons, 
we need not to restrict our scheme to quantum observables. It is applicable to any measurement design involving compatible and incompatible 
observables, see, e.g., \cite{BPS1}-\cite{EDSQ} for such experimental design in psychology. Here compatibility (incompatibility) is understood as the possibility (impossibility) 
of joint measurement and determination of jpd.   

\subsection{Terminology: observational, empirical, epistemic, and ontic}
\label{OEP}

This section can be useful for experts in quantum foundations. But, in principle,  one can skip it and jump directly to section \ref{RV}.

To be completely careful, in physics we should distinguish empirical probabilities obtained in experiments and theoretical probabilities given by the quantum formalism. The former are 
given by frequencies of outputs of observations. The von Mises frequency probability theory \cite{MI2, MI3} is the best formalism to handle them \cite{KHI}. The quantum theoretical probabilities
 are given by the Born rule. However, since 
the applicability of the quantum formalism was confirmed by numerous experiments, in physics we can identify quantum empirical and  theoretical probabilities.
(In principle, we should consider two sets of probabilities,  $p_{A_i B_j}^{\rm{exp}}$ and $p_{A_i B_j}^{\rm{QM}}.)$
 However, we want to present 
a very general framework covering even experiments outside physics, e.g., in psychology \cite{BPS1}-\cite{EDSQ}. 
Generally we interpret observables and corresponding probabilities  empirically.

The Bell type inequalities  cannot be derived in the observational framework (neither empirical nor theoretical). To derive them, one has to operate in the 
CP-framework. Here CP-theory is understood as the 
measure-theoretical approach  to probability proposed by  Kolmogorov  in 1933 \cite{K}. We remark that the Bell type inequalities 
cannot be derived  \cite{KHI} by using the von Mises frequency probability theory \cite{MI2, MI3}.  (In principle, this theory also can be considered as a CP-theory.)
 We also remark that von Neumann treated quantum probabilities in the von Mises framework \cite{VN}. So, the Bell argument 
is about comparison of the Hilbert space and measure-theoretic representations of probabilities.   

We also make a remark on the used terminology. In philosophy and quantum foundations, it is common to consider 
the {\it epistemic and ontic  descriptions} of natural (or mental) phenomena,   Atmanspacher and  Primas \cite{ATM}.
At the epistemic level we represent our knowledge about phenomena. The ontic level is related to ``reality as it is when nobody observes it.''

In this paper we speak about the {\it ``observational framework''}. 
This is more or less the same as the epistemic framework. But epistemic is typically related to a theorectical model. So, quantum mechanics
is an epistemic model. Our ``observational framework'' covers not only theoretical models, but even ``rough experimental data''.      

Although the term ``ontic'' is well established  in philosophy as well as in quantum foundations 
(and was widely used by one of the coauthors of this paper, e.g.,  \cite{KHRP2}), it seems that often the use of the notion of ``reality as it is''  
is really misleading. We can speak only about models and the ontic level of description is still our own (typically mathematical) description.  

We prefer to speak about  {\it ``counterfactual components
of a model''}. The ontic level of description is characterized by the presence of counterfactuals.

\subsection{Classical probability model  (BCHSH) for the Bohm-Bell experiment:  four random variables}
\label{RV}

Let $(\Lambda, {\cal F}, P)$ be some probability space \cite{K}. Here $\Lambda$ is the set of hidden variables (or in mathematics``elementary events''), 
${\cal F}$ is a $\sigma$-algebra of events, $P$  is a probability measure on ${\cal F}.$  

The notion of a $\sigma$-algebra can be disturbing for physicists. We remark that if $\Lambda$ is finite, then ${\cal F}$ is the collection of 
all its subsets.  In CP-modeling the CHSH framework it can be assumed  that $\Lambda$ is finite.

Consider two pairs of random variables $a_1, a_2: \Lambda \to \{\pm  1\}$ 
and $b_1, b_2: \Lambda \to \{\pm 1\}.$  These random variables are associated with observables $A_1, A_2, B_1, B_2.$ This is the Bell type  CP-model  for the observational  
framework  presented in section \ref{EPL}. Denote this CP-model  by ${\cal M}_{BCHSH}.$

We remark that the jpd  of four random variables $a_1, a_2, b_1, b_2$ is well defined:
 $$
P_{a_1 a_2 b_1 b_2}(\alpha_1, \alpha_2, \beta_1, \beta_2) 
$$
$$
=P(\lambda: a_1(\lambda)=\alpha_1,  a_2(\lambda)=\alpha_2,  b_1(\lambda)=\beta_1, 
  b_2(\lambda)=\beta_2),
$$
where $\alpha_i, \beta_j= \pm 1.$

In  model ${\cal M}_{BCHSH},$ one can form the CHSH linear combination of the correlations of the pairs of random variables $a_i, b_j$ 
\begin{equation}
\label{CHSH}
{\cal B}=  \langle  a_1  b_1\rangle  - \langle  a_1  b_2\rangle + \langle  a_2  b_1\rangle +\langle  a_2 b_2\rangle   
\end{equation}
and prove the CHSH-inequality:
\begin{equation}
\label{CHSH1}
\vert {\cal B} \vert \leq 2.
\end{equation}
Here 
\begin{equation}
\label{CHSHY}
\langle  a_i  b_j\rangle\equiv E(a_ib_j)= \int_\Lambda a_i (\lambda) b_j(\lambda) d P(\lambda)=\sum_{\alpha, \beta} \alpha \beta  P_{a_i b_j}( \alpha, \beta) .  
\end{equation}

We remark that probabilities for the joint measurements of $a$ and $b$ observables can be represented as the marginal probabilities 
for the quadruple  jpd, e.g.,   $P_{a_1 b_1} (\alpha, \beta) = \sum_{x, y} P_{a_1 a_2 b_1 b_2}(\alpha, x, \beta, y).$
This representatio plays the crucial role in the derivation of CHSH-inequality (\ref{CHSH1}). 
Moreover, by Fine's theorem\cite{Fine} the existence of the jpd is equivalent to satisfying
the CHSH-inequality.  

In principle, we can select $\Lambda$ as the set of vectors $\lambda=(\alpha_1, \alpha_2, \beta_1, \beta_2)$ with coordinates $\pm 1.$  Here probability $P$ is given by  
jpd; events are all possible subsets of this $\Lambda.$  

We remark that {\it model ${\cal M}_{BCHSH}$ contains counterfactual components,} e.g., the joint presence of the values 
of the incompatible observables, say $a_1(\lambda), a_2(\lambda).$ Consequently pair-wise jpds $P_{a_1 a_2}$ and $P_{b_1 b_2}$ as well quadrupole jpd 
$P_{a_1 a_2 b_1 b_2}$ are also counterfactual. By using the ontic-epistemic terminology we can say that {\it ${\cal M}_{BCHSH}$ is an ontic model for 
the epistemic model - quantum mechanics.}

Now consider the observational probabilities $p_{A_i B_j}.$ The BCHSH-coupling between the observational and CP descriptions  is straightforward, it will be presented in the next section.

\subsection{BCHSH-rule for  correspondence between observational and classical probabilities} 
\label{CC1}

The observational framework (section \ref{EPL}) is coupled with CP-model ${\cal M}_{BCHSH}$ by the following correspondence rule: 

\medskip

{\it The observational probabilities   $p_{A_i B_j}$ are identified with the CP-probabilities $P_{a_i b_j}.$ }

\medskip

This coupling leads to contradiction, because the CHSH linear combination composed of observational  correlations (either experimental or quantum theoretical):
\begin{equation}
\label{CHSHq}
{\cal B}_{\rm{observational}}=  \langle  A_1  B_1\rangle  - \langle  A_1  B_2\rangle + \langle  A_2  B_1\rangle +\langle  A_2 B_2\rangle   
\end{equation}
can violate  CHSH-inequality  (\ref{CHSH1}); generally 
\begin{equation}
\label{CHSH1tt}
\vert {\cal B}_{\rm{{observational}}}\vert > 2.
\end{equation}
One can conclude that  {\it CP-model ${\cal M}_{BCHSH}$ is not adequate neither to the quantum (epistemic) model nor to the experimental situation.}

This mismatching related to concrete CP-model ${\cal M}_{BCHSH}$ and the BCHSH correspondence rule is commonly interpreted too generally:{\it  as the impossibility of the CP-description 
of quantum phenomena, impossibility to represent quantum states by probability measures and  
quantum observables (generally incompatible) by classical random  variables.}

We remark that typically  physicists speak about realism and locality. In the CP-framework, realism is encoded in the functional representation of observables by random variables, locality 
(noncontextuality) is encoded in single indexing of random variables, say $a_i$ is indexed by solely its experimental setting $i$ (see \cite{D7}).   For PBS, this is the concrete 
orientation angle $\theta_i.$ Thus the statement about mismatching of model   ${\cal M}_{BCHSH}$ and quantum mechanics is typically stated as mismatching between
local realism and quantum mechanics.

\subsection{Missed component of experimental arrangement}

In the CHSH observational framework, the correlations composing quantity  ${\cal B}_{\rm{observational}}$ cannot be measured jointly. The concrete experiment can be performed only for one 
fixed pair of indexes $(i, j),$ experimental settings (orientations of PBSs). Generally these settings are selected randomly, by using two random generators $R_A$ and $R_B$ 
taking values $1,2.$ What 
are the theoretical counterparts of these random generators in  ${\cal M}_{BCHSH}?$  They are absent. So, CP-model ${\cal M}_{BCHSH}$ is inadequate the observational 
framework.  One sort of randomness, namely, generated by $R_A, R_B$ is missed. We shall present another CP-model 
corresponding the real experimental situation: the observational BCHSH-framework (section \ref{EPL}) with supplementary observables $R_A, R_B.$

By proceeding in this way we follow the Copenhagen interpretation of quantum mechanics. Bohr always emphasized: {\it all components of the 
experimental arrangement (context) have to be taken into account} \cite{BR}.
Experimenters strictly follow the Copenhagen interpretation. Random generators play the fundamental role in the experiments of Bohm-Bell type. However, these generators
are absent in CP-model ${\cal M}_{BCHSH}.$  

\section{Bohm-Bell  type experiments: taking into account random generators}

At the observational level, we plan to complete the standard description of the Bohm-Bell type experiments (section \ref{EPL}) by taking into account 
the aforementioned ``missed components of the 
experimental  arrangement''. Then we shall consuruct a CP-model which will be adequate to the completed  observational framework. It will 
take into account ``missed component of randomness''.  Denote such a CP-model under construction by ${\cal M}_{\rm{KH}}.$

\subsection{Description of  (six) observables}
\label{EPL1}

Following Bohr, we treat random generators $R_A$ and $R_B$  as a part of  experimental arrangement. Instead of the observational framework with four observables (section \ref{EPL})
$A_1, A_2, B_1, B_2,$ we consider the framework with six  observables $A_1, A_2, B_1, B_2, R_A, R_B.$ The latter two observables are compatible, i.e., they can be jointly measurable; 
moreover, they are compatible with each of four ``basic observables''  $A_1, A_2, B_1, B_2$ (see \cite{KHRF} for the mathematical representation of these six observables 
within the quantum operator formalism).  
In principle, in the real experimental situation one can assume that observables $R_A$ and $R_B$ are independent.  For for the moment, we proceed without this assumption. 

To improve the visibility of the role of random generators, in physics  we can consider 
the experimental design of the pioneer experiment performed by Aspect, see \cite{AS}.  In the modern experimental design, there are 
 two beam splitters, one on the $A$-side and another on the $B$-side, and two devices for random selection of orientations  on the corresponding sides. 
Aspect considered four beam splitters and two switchers preceding corresponding  pairs of beam splitters. The $A$-switcher selects randomly one of the beam splitters 
on the $A$-side; the $B$-switcher selects randomly one of the beam splitters on the $B$-side (switchers open optical channels to corresponding beam splitters).
For this design, it is natural to introduce the additional value of observables, we set $A_i=0$ ($B_j=0)$ if its input channel is closed by the random switcher. 

We consider the ideal experiment with 100 \% of efficiency of the whole experimental scheme, i.e., including detector, beam splitters, an optical fibers.

\subsection{Complete CP-model:  six random variables}
\label{RV1}

Let again $(\Lambda, {\cal F}, P)$ be some probability space. 
We want to introduce  random variables $a_1, a_2, b_1, b_2$  associated with observables $A_1, A_2, B_1, B_2,$ but not so straightforwardly as in ${\cal M}_{\rm{BCHSH}}.$
Additionally, we consider two random variables  $r_A, r_B: \Lambda \to \{1,2\}$ associated with the random generators. 
 Besides of values $\pm 1,$ random variables $a_1, a_2, b_1, b_2$ can take the value zero. 

The zero-value is determined by governing selections of measurement settings, i.e., $A_1, A_2, B_1, B_2,$ 
by random generators $R_A$ and $R_B.$ In our CP-model, it has the form: 
\begin{itemize}
\item $a_i=0$  (with probability one), if the $i$-setting was not selected, i.e., $r_A\not=i;$ 
\item  $b_j=0$ (with probability one),  if the $j$-setting was not selected, i.e., $r_B\not=j.$ 
\end{itemize}
We remark that in our model the zero-value has nothing to do with detection's inefficiency(as is often considered in modeling the Bohm-Bell experiment). We model the experimental situation 
with detectors having 100\% efficiency. 

\subsection{Constraints on joint probabilities implied by matching condition}

In  terms of probability the condition of  $a-r_a$ matching can be written as follows:
\begin{equation}
\label{MA1}
P(a_i=0 \vert r_a=j)=1, i \not= j.
\end{equation}
It implies that   
\begin{equation}
\label{MA1b}
P(a_i=\alpha \vert r_a=j)=0, \alpha= \pm 1, \; i \not= j.  
\end{equation}
Thus RV $a_i$ cannot take values $\pm 1$ if $r_a \not=i.$ This is the CP-presentation of the impossibility 
to measure observable $A_i$ if  random generator $R_A\not=i.$
Equality  (\ref{MA1b}) implies
\begin{equation}
\label{e528}
P(a_i=\alpha, r_a=j)=0, \alpha= \pm 1, i \not= j.
\end{equation}

 In the same way,  the condition of  $b-r_a$ matching can be written as follows: 
\begin{equation}
\label{MA2}
P(b_i=0 \vert r_b=j)=1, i \not= j.
\end{equation}
 This condition implies 
\begin{equation}
\label{e529}
P(b_i=\beta,  r_b=j)=0, \beta= \pm 1, i \not= j.
\end{equation}

From equalities (\ref{MA1}),  (\ref{MA2}), we obtain   
\begin{equation}
\label{MA3}
P(a_i=0,  r_a=j)=P(r_a=j),  \; P(b_i=0,  r_b=j)=P(r_b=j),  i \not= j. 
\end{equation}
In turn, these equalities imply
\begin{equation}
\label{MA4}
P(a_i=0,  r_a=i)=P(r_a=i),  \; P(b_i=0,  r_b=i)=P(r_b=i). 
\end{equation}

The jpd  of six random variables $a_1, a_2, b_1, b_2, r_A, r_B$ is well defined:
 $$
P_{a_1 a_2 b_1 b_2 r_a r_b}(\alpha_1, \alpha_2, \beta_1, \beta_2, \gamma_1, \gamma_2) 
$$
$$
=P(\lambda: a_1(\lambda)=\alpha_1,  a_2(\lambda)=\alpha_2,  b_1(\lambda)=\beta_1, 
  b_2(\lambda)=\beta_2,   r_A(\lambda)=\gamma_1,  r_B(\lambda)=\gamma_2),
$$
where $\alpha_i, \beta_j= 0, \pm 1, \gamma_k=1,2.$

The matching condition implies that, e.g.,  $P_{a_1 a_2 b_1 b_2 r_a r_b}(\alpha_1, \pm 1, \beta_1, \beta_2, 1, \gamma_2)=0.$   
Thus only 16 components of the jpd are different from zero:  
$$
P_{a_1 a_2 b_1 b_2}(\alpha, 0, \beta, 0, 1, 1),  P_{a_1 a_2 b_1 b_2}(\alpha, 0, 0 , \beta,  1, 2), 
$$
$$
P_{a_1 a_2 b_1 b_2}(0, \alpha,  \beta, 0,  2, 1),  P_{a_1 a_2 b_1 b_2}(0, \alpha, 0 , \beta, 2, 2), 
$$
where $\alpha, \beta= \pm 1.$ 

\subsection{Correspondence between observational and classical conditional  probabilities}
\label{CC2}

Now consider the observational probabilities $p_{A_i, B_j}.$ These are probabilities for the fixed pair of experimental settings $(i, j).$ Their counterparts in 
 CP-model ${\cal M}_{\rm{KH}}$ are obtained by conditioning on the fixed values of random variables $r_A$ and $r_B.$ The rule of correspondence  
between observational and CP-probabilities is based on the following identification: 
\begin{equation}
\label{e3}
p_{A_i B_j} (\alpha, \beta) = P(a_i= \alpha, b_j= \beta\vert r_A= i, r_B=j),
\end{equation}
where $\alpha, \beta= \pm 1.$  Thus
\begin{equation}
\label{e3a}
p_{A_i B_j} (\alpha, \beta) =  \frac{P(\lambda \in \Lambda: a_i (\lambda)= \alpha, b_j (\lambda)= \beta,  r_A(\lambda)= i, r_B(\lambda)=j)}{P(\lambda \in \Lambda: r_A(\lambda )= i, r_B(\lambda )=j)}.
\end{equation}

This correspondence rule for the ``basic observables'' is completed by the similar rule for random generators $R_A$ and $R_B:$
\begin{equation}
\label{e3z}
p_{R_A R_B} (i, j) = P(\lambda \in \Lambda: r_a(\lambda )=i,  r_b(\lambda) =j\}.
\end{equation}

\subsection{Violation of the CHSH-inequality by conditional correlations}

Conditioning  on the selection of experimental settings plays the crucial role. The  CP-correlations are based on the conditional probabilities
\begin{equation}
\label{e5}
\Big \langle  a_i  b_j  \Big\rangle  \equiv  E(a_i b_j\vert r_A=i,  r_B=j) 
\end{equation}
$$
= \sum_{\alpha, \beta= \pm 1} \alpha \beta P(a_i= \alpha, b_j= \beta\vert r_A= i, r_B=j).
$$
 We can form the CHSH linear combination of conditional correlations of RVs:   
\begin{equation}
\label{e5z}
\tilde{\cal B}= \Big \langle  a_1  b_1  \Big\rangle - \Big \langle  a_1  b_2  \Big\rangle +\Big \langle  a_2  b_2  \Big\rangle + \Big \langle  a_2  b_2  \Big\rangle
\end{equation}
It is possible to find such classical probability spaces that 
$$
\vert \tilde{\cal B} \vert > 2.
$$
Since each conditional probability is also a probability measure and since RVs $a_i, b_j$ take values in [-1, +1], the conditional expectations 
$E(a_i b_j\vert r_A=i,  r_B=j)$ are bounded by 1, so 
$$
\vert \tilde{\cal B}\vert \leq 4.
$$
Thus the common claim on mismatching of the CP-description with  quantum mechanics and experimental data was not justified.

In principle, one can consider linear combination ${\cal B}$ composed of correlations $\langle  a_1  b_1  \rangle$ which are not conditioned on selection of experimental settings.
Such ${\cal B}$ satisfies the CHSH-inequality.  But such correlations cannot be identified with experimental ones.

\subsection{Construction of jpd from observational probabilities}

Correspondence rules (\ref{e3}),  (\ref{e3z}) imply
\begin{equation}
\label{e6}
P_{a_1 a_2 b_1 b_2 r_a r_b}(\alpha_1, \alpha_2, \beta_1, \beta_2, i, j) 
= p_{A_i B_j} (\alpha, \beta) p_{R_A R_B}(i, j) , \alpha, \beta=\pm 1, 
\end{equation}

From this equality, we can determine all nonzero components the jpd:
$$
p(\alpha, 0, \beta, 0, 1, 1)=  p_{A_1 B_1} (\alpha, \beta) p_{R_A R_B}(1, 1), 
p(\alpha, 0, 0 , \beta,  1, 2)= p_{A_1 B_2} (\alpha, \beta) p_{R_A R_B}(1, 2), 
$$
$$
p(0, \alpha,  \beta, 0,  2, 1) =p_{A_2 B_1} (\alpha, \beta) p_{R_A R_B}(2, 1), 
 p(0, \alpha, 0 , \beta, 2, 2)= p_{A_2 B_2} (\alpha, \beta) p_{R_A R_B}(2, 2)
$$
In model ${\cal M}_{\rm{KH}},$ the jpd is completely determined by observational (epistemic) probabilities. In contrast to ${\cal M}_{\rm{CHSH}},$ there 
are no counterfactual probabilities.

\section{(No-)signaling}

\subsection{No-signaling in quantum physics}

In the observational framework for the Bohm-Bell type experiment,   the condition of no-signaling 
is formulated in the probabilistic terms. There is no-signaling, from the $B$-side to the $A$-side, if the $A$-marginals of jpds 
$p_{A_i B_j}$    
\begin{equation}
\label{e5ztt}
M_{ij}(\alpha) = \sum_{\beta=\pm 1} p_{A_i B_j} (\alpha, \beta), \; i=1,2,
\end{equation}
do not depend on the index $j.$ 

This notion of signaling need not be rigidly coupled to quantum observables. It can be applied to any measurement design in that $A_i$ is compatible with both $B_j, 
j=1,2,$ but $B_1$ and $B_2$ are incompatible, i.e., we are not able to perform their joint measurement. No-signaling from the $A$-side to the $B$-side is defined 
in the same way.

In physics, signaling is often understood as real signaling from the $B$-side to the $A$-side and even, what is worse, 
from the $B$ system to the $A$-system. By constructing the CP-model, we can clarify the meaning of (no-)signaling at the level of RVs and then observations.

\subsection{No-signaling as condition of independence of random variables}

Now we proceed with CP-model ${\cal M}_{\rm{KH}}.$ Let us fix $r_a=i.$ For any value $r_b=j,$ consider conditional $a_i$-marginal
\begin{equation}
\label{KTT}
m_{ij} (\alpha)= \sum_\beta P(a_i=\alpha, b_j=\beta \vert r_a= i, r_b=j), i=1,2.
\end{equation}
By correspondence rules (\ref{e3}),  (\ref{e3z}) 
\begin{equation}
\label{KTT}
M_{ij}(\alpha)= m_{ij}(\alpha). 
\end{equation}
The marginal  $m_{ij}(\alpha)$ does not depend on the $j$-settings governed by $r_b$ under the following assumption:  

\medskip
${\bf I_{a_i}}$ {\it The pair of RVs $a_i,   r_a$ does not depend on RV $r_b.$}       
\medskip

Under this  assumption 
\begin{equation}
\label{KTT1}
m_{ij}(\alpha) = P(a_i=\alpha \vert r_a = i).
\end{equation}
This is the   conditional-probability  version of no-signaling for $a_i.$ 
To prove equality (\ref{KTT1}), we first remark 
\begin{equation}
\label{KTT2}
m_{ij}(\alpha)  = P(a_i=\alpha \vert r_a= i, r_b=j)
\end{equation}
(since the conditional probability is a probability measure).
Hence, 
\begin{equation}
\label{KTT2}
m_{ij}(\alpha)  =  \frac{P(a_i=\alpha,  r_a= i, r_b=j)}{P(r_a= i, r_b=j)}= \frac{P(a_i=\alpha,  r_a= i) P (r_b=j)}{P(r_a= i) P(r_b=j)}  
\end{equation}
and this proves (\ref{KTT1}). 

Now, let us assume that RVs $r_a$ and $r_b$ are independent. (From the experimental viewpoint, this is the very natural assumption.)
Suppose that, for   $\alpha= \pm1,$ the marginal $m_{ij} (\alpha)$ does not depend on $j.$
Generally this marginal can be represented in the form:  
\begin{equation}
\label{KTT}
m_{ij} (\alpha)= P(a_i=\alpha \vert r_a= i, r_b=j)= \frac{P(a_i=\alpha, r_a= i \vert  r_b=j)}{P(r_a= i\vert r_b=j)}=  \frac{P(a_i=\alpha, r_a= i \vert  r_b=j)}{P(r_a= i)}.
\end{equation}
The right-hand side does not depend on $j$ only if $P(a_i=\alpha, r_a= i \vert  r_b=j)= P(a_i=\alpha, r_a= i )$    (see appendix). This is the condition of independence of the pair of RVs
$a_i, r_a$ from RV $r_b$.

In the same way,  consider the assumption 

\medskip

${\bf I_{b_j}}$ {\it   The pair of random variables $b_j,   r_b$ does not depend on $r_a.$ } 

\medskip

Under this assumption
$$
m_{ij}(\beta)= \sum_\alpha P(a_i=\alpha, b_j=\beta \vert r_a= i, r_b=j) =  P(b_j=\beta \vert  r_b=j).
$$
This is the conditional version of no-signaling for random variable $b_j.$

The  CP-presentation of  no-signaling in terms of  conditional probabilities, see ${\bf I_{a}}, {\bf I_{b}},$ explains the meaning of signaling. 
For example, $b \to a$ signaling means   either interdependence of random generators $r_a$ and $r_b,$ or dependence of $a$-RVs on random generator $r_b.$

Under the assumption of independence of RVs $r_a$ and $r_b$ representing the random generators,  
$b \to a$  signaling  has the meaning of dependence of $a$-variables on random generator $r_b,$ i.e., 
the latter governs not only $b$-variables, but even the $a$-variables.  

\subsection{Interpretation of no-signaling: from random variables to  observables}

By using (\ref{KTT})  we can lift the CP-interpretation of no-signaling to the level of observables. 
Let us consider the case of independent random generators $R_A$ and  $R_B$ represented by independent RVs $r_a$ and $r_b.$
The absence of $B\to A$ signaling for observables, i.e., independence $M_{ij}(\alpha)$  from index $j,$ is equivalent  
to the absence of $b\to a$ signaling  RVs. Hence, {\it at the observational level  $B\to A$  no-signaling  has the meaning of independence of $A$-observables 
from selection of experimental settings governed by random generator $R_B.$} 

We stress that ${\cal M}_{\rm{KH}}$ can serve as a  CP-model for quantum probabilities, i.e., probabilities described by the quantum formalism with the aid of the Born rule.
Thus the absence of signaling in the quantum description of the Bohm-Bell experiment has very natural CP-explanation: selection of $A$-settings depends only on the random generator 
$R_A$ and selection of $B$-settings depends only on the random generator 
$R_B.$ 

\subsection{(No-) signaling in experiments in quantum physics and psychology}

In quantum physics the problem of the presence of  signaling patterns in statistical  data collected in  the Bohm-Bell type experiments was highlighted   in the work \cite{AD1} 
(it seems, it was the first paper on this problem). Since the quantum formalism predicts the absence of signaling, such signaling patterns were considered as
a consequence of the improper experimental performance. After the pioneer paper  \cite{AD1}, experimenters started to pay attention to signaling. Tremendous  efforts of experimenters 
to eliminate technicalities which may lead to signaling were culminated in the breakdown experiments of Vienna's group \cite{B2} and NIST's group \cite{B3}. (Unfortunately, the first 
experiment claiming to be loophole free \cite{B1} suffers of strong signaling, see \cite{AD2}).

As was found by Dzhafarov and the coauthors, see, e.g., \cite{D3, D6}, 
the psychological experiments of the Bohm-Bell type generated statistical data with statistically non-negligible signaling patters.
(These are experiments to test quantum contextuality in the psychological analogs of the Bell-Bohm type experiments
\cite{BPS1}-\cite{EDSQ}. So, the issue of nonlocality is not involved.) In psychology we do not have theoretical justification of the absence of signaling.
Therefore it is not clear whether the mental signaling is a consequence of improper experimental design and 
performance or this is the fundamental  feature of experiments with humans.

\section{Concluding remarks}

We presented the brief review on CP-representations of quantum probability. Then the paper was concentrated on one special representation based on the {\it conditional 
probability interpretation} of quantum probabilities \cite{KHRB2}. The formalism of the latter article was described in the very general framework 
covering the experimental schemes  of the Bohm-Bell type. Such experimental schemes need not be coupled to quantum physics. In particular, they can be realized for experiments
with humans. As was found by Dzhafarov and the couathors, the latter experiments are characterized by the presence of statistically significant signaling patterns. In this paper, 
we analyzed the CP-meaning of signaling in the conditional probabilistic model. We found that signaling can be described as simply dependence of random variables.     

We highlight the basic impacts of the CP-representation of the experimental schemes of the Bohm-Bell type:
\begin{enumerate}
\item It demystifies quantum probability theory - representation of probabilities by complex amplitudes and observables by Hermitian operators:
\item It justifies the use of  CP-based  mathematical statistics for analysis of data from quantum experiments.\footnote{For example, to check statistical significance of a violation 
of a Bell type inequality, experimenters always use classical mathematical statistics, e.g., $p$-values or Chebyshov inequality \cite{W}. However, by demonstrating that a violation 
of this Bell type inequality is statistically significant, one has to understand that the standard CP-representation  based on model  ${\cal M}_{BCHSH}$ is impossible. Therefore the preceding 
CP-based statistical analysis justifying the hypothesis on the violation of the Bell type inequality was meaningless. Of course, one can appeal to quantum theory of decision making. But suh 
appealing is meaningless in comparing classical and quantum descriptions. In contrast, with CP-models  ${\cal M}_{KH}$ or  ${\cal M}_{DZ}$ one can proceed with CP-based 
statistics. Of course, these models are different both from the foundational and technical viewpoints. Analysis of data with the aid of ${\cal M}_{KH}$ can be used to justify statistical
significance of violation of the CHSH-inequality for experimental probabilities which are interpreted as classical probabilities conditional on selection of experimental settings.}
\item It clarifies the meaning of (no-)signaling as independence-dependence of classical random variables. 
\end{enumerate}
 
 Finally, we emphasize once again the foundational impact of Ballentine's works \cite{BL}-\cite{BL2} on the conditional probabilistic interpretation of quantum probabilities.  These works 
 stimulated development of contextual probability theory \cite{KHRC}. As was found in \cite{KHRB2}, the quantum contextual probabilities generated in experiments of the 
Bohm-Bell type can be even represented as classical probabilities (see also  \cite{D1}-\cite{D7}).

 \section*{Appendix}
 
 Consider two RVs $X$ and $Y$ Here $X$ is an arbitrary discrete RV, $X=x_1,..., x_m,$ and $Y$  is a dichotomous RV, $Y=1,2.$ Suppose that, for each $x,$ conditional probability  
 $P(X=x \vert Y=j)$ does not depend no $j.$ We want to show that this implies that, in fact,   
\begin{equation}
\label{lo}
P(X=x \vert Y=j)=P(X=x),
\end{equation}
 i.e., that RVs $X$ and $Y$ are independent. 

Set $A_x=\{\lambda \in \Lambda: X(\lambda)=x\}$ and $B_j= \{\lambda \in \Lambda: Y(\lambda)=j\}.$ We have 
$$
P(A_x\vert B_1)= P(A_x\vert B_2), \; \mbox{i.e.}\; P(A_x \cap  B_1)= \frac{P(B_1)}{P(B_2)} P(A_x \cap  B_2),
$$
or
$$
 P(A_x \cap  B_1)= \frac{P(B_1)}{P(B_2)}\Big[P(A_x) - P(A_x \cap  B_1)\Big],
$$
i.e.
$$  
P(A_x \cap  B_1) \Big[1 + \frac{P(B_1)}{P(B_2)}\Big] = \frac{P(B_1)}{P(B_2)}P(A_x).
 $$
 Thus we obtained 
 $$
 P(A_x \cap  B_1)= P(B_1) P(A_x)
 $$
 This also implies that  $P(A_x \cap  B_2)= P(B_2) P(A_x).$ Hence, equality (\ref{lo}) holds and RVs $X$ amd $Y$ are independent.  
 
 \section*{Acknowledgments}

 This work was financially supported by Government of Russian Federation, Grant 08-08 and by Ministry of Education and 
 Science of the Russian Federation within the Federal Program "Research and development in priority areas for the development
 of the scientific and technological complex of Russia for 2014-2020", Activity 1.1, Agreement on Grant No. 14.572.21.0008 of
 23 October, 2017, unique identifier: RFMEFI57217X0008.


\begin{thebibliography}{99}
 
 \bibitem{W} Wigner, E. On the quantum correction for thermodynamic equilibrium. {\it Phys. Rev.} {\bf 1932}, {\it 40}, 749–759.
 
 \bibitem{H} Husimi, K. Some formal properties of the density matrix. {\it  Proc. Phys. Math. Soc. Jpn.} {\bf  1940}, 22, 264–314.

 \bibitem{K}  Kano, Y. A new phase-space distribution function in the statistical theory of the electromagnetic field.
{\it  J. Math. Phys.} {\bf 1965}, 6, 1913–1915. 
 
\bibitem{G} Glauber, R.J. Coherent and incoherent states of the radiation field. {\it  Phys. Rev.} {\bf  1963}, {\it  131}, 2766–2788. 
 
\bibitem{S} Sudarshan, E.C.G. Equivalence of semiclassical and quantum mechanical descriptions of statistical light
beams. {\it  Phys. Rev. Lett.} {\bf  1963}, {\it  10}, 277–279.

\bibitem{Brida} Brida, G.;  Genovese, M.;  Gramegna, M.; Novero, C.;   Predazzi, E. A first test of Wigner function local realistic
model. {\it Phys. Lett}  A  {\bf 2002}, {\it 299}, 121.

\bibitem{MA1} Mancini, S.; Man'ko, V.I.; Tombesi, P. Symplectic tomography as classical approach to quantum systems.
{\it Phys. Lett.} A {\bf  1996}, {\it 213}, 1–6.

\bibitem{MA2} Dodonov, V.V.; Man'ko, V.I. Positive distribution description for spin states. {\it Phys. Lett.} A {\bf  1997}, {\it 229}, 335–339.

\bibitem{MA3} Man'ko, V.I.; Man'ko, O.V. Spin state tomography. {\it J. Exp. Theor. Phys.} {\bf  1997}, {\it 85}, 430–434. 

\bibitem{KHRP0a} Khrennikov, A. Prequantum classical statistical field theory: Schr\"odinger dynamics of entangled 
systems as a classical stochastic process. {\it Found. Physics} {\bf 2011}, {\it  41},  317-329.

\bibitem{KHRP0} Khrennikov, A. Towards a field model of prequantum reality. {\it Found.  Phys.} {\bf 2012}, {\it  42},  725-741.   

\bibitem{KHRP1} Khrennikov, A.  {\it Beyond Quantum};    Pan Stanford Publishing:  Singapore,  2014.

\bibitem{KHRP2}  Khrennikov, A.   Quantum epistemology from subquantum ontology: Quantum mechanics from theory of
classical random fields. {\it Annals of Physics} {\bf 2017}, {\it  377},  147-163.

\bibitem{KHRP3a} Khrennikov, A. Quantum probabilities and violation of CHSH-inequality from classical random signals 
and threshold type detection scheme. {\it Prog. Theor. Phys.} {\bf 2012}, {\it  128},   31-58.

\bibitem{KHRP3b} Khrennikov, A.; Nilsson, B. ; Nordebo, S.   On an experimental test of prequantum theory of classical random fields : an estimate from above of 
the coefficient of second-order coherence. {\it Int. J. Quantum Inf.} {\bf 2012},  {\it 10},  Article ID: 1241014.

\bibitem{K} Kolmogorov, A. N.  \textit{Foundations of the theory of probability}. Chelsea Publishing Company: New York, 1956.

\bibitem{VN}  Von Neumann, J. {\it  Mathematical Foundations of Quantum Mechanics.} Princeton University Press, Princeton, 1955.

\bibitem{BE2} Bell, J. S. \textit{Speakable and Unspeakable in Quantum Mechanics}; \emph{\ }Second Edition,  Cambridge Univ. Press: Cambridge, 2004.

\bibitem{BE1} Bell, J.S. On the Einstein Podolsky Rosen paradox. \textit{Physics}\emph{\ }\textbf{1964}\emph{, }\textit{1,} 195--200.


\bibitem{BC1} Kupczynski, M. Can Einstein with Bohr debate on quantum mechanics be closed? \textit{Phil. Trans. Royal Soc.} A \textbf{2017}, 
\textit{375,} N 2106, 2016039.

\bibitem{BC1a} Kupczynski, M. Closing the door on quantum nonlocality. {\it Entropy} {\bf  2018}, {\it 20}, 877.

\bibitem{BC2} Khrennikov, A. After Bell. \textit{Fortschritte der Physik (Progress in Physics)} \textbf{2017}, \textit{65}, N 6--8, 1600014.

\bibitem{BC3} Khrennikov, A. Bohr against Bell: complementarity versus nonlocality. \textit{Open Physics}\emph{\ }\textbf{2017}\emph{, }\textit{15,}
734--738.


\bibitem{BC4} De Raedt, H.;  Katsnelson, M. I.; Michielsen, K.   Logical inference derivation of the quantum theoretical description of Stern–Gerlach 
and Einstein–Podolsky–Rosen–Bohm experiments. {\it Annals of Physics} {\bf 2018}  {\it 396},   96-118.

\bibitem{L8} Khrennikov, A.; Basieva, I. Towards experiments to test violation of the original Bell inequality. \textit{Entropy} \textbf{2018}, 
\textit{20,} 280:1--280:12.

\bibitem{L9} Khrennikov, A.; Loubnets, E. Evaluating the maximal  violation of the original Bell inequality by two-qudit states exhibiting 
perfect correlations/anticorrelations. {\it Entropy} {\bf 2018}, {\it 20}, 829.


\bibitem{CHSH} Clauser, J. F.; Horne, M. A.; Shimony, A and Holt, R. A. Proposed experiment to test local hidden-variable theories. \textit{Phys.
Rev. Lett.} \textbf{1969,} \emph{23} (15) 880--884.


\bibitem{Fine1} Fine, A. Hidden Variables, Joint Probability, and the Bell Inequalities. \textit{Phys. Rev. Lett.}\emph{\ }\textbf{1982,}\emph{\ }%
\textit{48,} 291.

\bibitem{Fine2} Fine, A.  (1982). Joint distributions, quantum correlations, and commuting observables. {\it Journal of Mathematical
 Physics, 23}, 1306.

\bibitem{KHRB1} Avis, D.;  Fischer, P.;  Hilbert, A.;  Khrennikov, A. Single, Complete, Probability Spaces Consistent With EPR-Bohm-Bell Experimental Data.
In: {\it  Foundations of Probability and Physics-5}, AIP Conference Proceedings, {\it 1101}, 294-301, 2009.

\bibitem{KHRB2}  Khrennikov, A.   CHSH inequality: quantum probabilities as classical conditional probabilities. 
{\it Foundations of Physics} {\bf 2015}, {\it  45},   711-725.   

\bibitem{D1} Dzhafarov, E. N.; Kujala, J. V. Selectivity in probabilistic causality: Where psychology runs into quantum physics.
{\it J. Math. Psych.} {\bf 2012}, {\it  56}, 54-63.

\bibitem{D3} Dzhafarov, E. N.;  Zhang, R.;  Kujala, J.V.  Is there contextuality in behavioral and social systems?
{\it Phil. Trans.  Royal Soc. A} {\bf 2015}, {\it  374},  20150099.

\bibitem{D4} Dzhafarov, E. N.;   Kujala, J. V.   Probabilistic contextuality in EPR/Bohm-type systems with signaling allowed.
In E.  Dzhafarov,  S.  Jordan,  R.  Zhang,  \&  V. Cervantes  (Eds.), {\it  Contextuality from Quantum Physics to Psychology}, 
pp. 287-308.  WSP: New Jersey, 2015 

\bibitem{D5} Dzhafarov, E. N.;  Kujala, J. V.  Context-content systems of random variables: The contextuality-by default
theory. {\it J, Math. Psych.} {\bf 2016}, {\bf 74}, 11-33.

\bibitem{D6} Dzhafarov, E. N.;  Kujala, J. V.;  Cervantes, V. H.;  Zhang, R.;  Jones, M.  
On contextuality in behavioral data. {\it Phil. Trans. Royal Soc.} A {\bf 2016}, {\it 374},  20150234.

\bibitem{D7a} Dzhafarov, E.N.;  Kujala, J.V.  Contextuality analysis of the double slit experiment (with a glimpse into three slits).
{\it Entropy} {\bf  2018}, {\it 20}, 278.

\bibitem{D7} Dzhafarov, E. N.;  Kon, M. On universality of classical probability with contextually labeled random variables. 
{\it  J. Math. Psych.}  {\bf 2018}, {\it 85}, 17-24.

\bibitem{BL} Ballentine, L.E. The statistical interpretation of quantum mechanics. {\it Rev. Mod. Phys.} {\bf  1989}, {\it 42}, 358--381

\bibitem{BL1} Ballentine, L.E.  {\it Quantum Mechanics: A Modern Development}; WSP: Singapore, 1998

\bibitem{BL2} Ballentine, L. E.:  Interpretations of probability and quantum theory. In:  Khrennikov, A. Yu. (ed)  Foundations of
Probability and Physics, {\it Quantum Probability and White Noise Analysis} {\it  13},  71-84. WSP: Singapore, 2001.

\bibitem{BR} Bohr, N.   {\it The philosophical writings of Niels Bohr}, 3 vols. Ox Bow Press: Woodbridge, Conn., 1987.

\bibitem{PL} Plotnitsky, A.  {\it Niels Bohr and Complementarity: An Introduction.} Springer: Berlin and New York, 2012.

\bibitem{OSID} Khrennikov, A. Bell could become the Copernicus of probability. {\it Open Systems and Information Dynamics} {\bf 2016}, {\it   23},  1650008.

 \bibitem{KHRC1}  Khrennikov, A.  Schr\"odinger dynamics as the Hilbert space projection of a realistic contextual probabilistic dynamics.
{\it Europhys. Lett.}  {\bf 2005} {\bf 69}, 678-684.

\bibitem{KHRC2}  Khrennikov, A. Algorithm for quantum-like representation: Transformation of probabilistic data into vectors
on Bloch's sphere. {\it Open Systems and Information Dynamics}, {\bf 2008}, {\it 15}, 223-230.

\bibitem{KHRC}  Khrennikov,  A. {\it Contextual approach to quantum formalism,} Springer: Berlin-Heidelberg-New York, 2009.

 \bibitem{BPS1} Conte, E.; Khrennikov, A.; Todarello, O.; Federici. A.; Mendolicchio, L.;    Zbilut, J. P.   A preliminary experimental verification on the possibility of Bell inequality violation in mental states. 
{\it  NeuroQuantology} {\bf 2008}, {\it  6}, 214-221.

\bibitem{BPS2} Asano, M.;  Khrennikov, A.;   Ohya, O.;   Tanaka, Y.;   Yamato,  I.    \textit{Quantum Adaptivity in Biology: from Genetics to Cognition.}  Springer: Heidelberg-Berlin-New York, 2015.

\bibitem{EDSQa}  Dzhafarov, E. N.;  Kujala, J. V.  Snow queen is evil and beautiful: experimental evidence for probabilistic 
contextuality in human choices.      {\it  J. Math. Psych.} {\bf  2018},   {\it 85}, 17-24.

\bibitem{EDSQ} Platonov A.V.;  Poleshchuk E.A.;  Bessmertny I.A.;  Gafurov N.R. 
Using quantum mechanical framework for language modeling and information retrieval.
12th IEEE International Conference on Application of Information and Communication Technologies, AICT 2018 - Conference 
Proceedings (Almaty, Kazakhstan 17-19 Oct 2018) , 99-102, 2018.

\bibitem{MA4} Man'ko, M. A.;   Man'ko,  V. I.  New entropic inequalities and hidden correlations in quantum suprematism pictue of qudit states.
{\it Entropy} {\bf  2018}, {\it 20}, 692.

\bibitem{MI2} Von Mises, R. {\it Probability, Statistics and Truth. Macmillan};  London, 1957.

\bibitem{MI3} Von Mises, R.  {\it The Mathematical Theory of Probability and Statistics}; Academic: London, 1964.

\bibitem{KHI} Khrennikov A. {\it Interpretations of Probability}; VSP Int. Sc. Publ.: Utrecht (1999);
the second edition (corrected and completed), De Gruyter: Berlin, 2009.

\bibitem{ATM} Atmanspacher, H.;  Primas, H.  Epistemic and ontic quantum realities. In:  Adenier, G.,  Khrennikov, A. Yu. (eds) {\it Foundations
of Probability and Physics-3},   {\it 750}, 49-62. AIP: Melville, NY, 2005.

\bibitem{KHRF} Khrennikov, A. 
Unconditional quantum correlations do not violate Bell's inequality. {\it Found. Phys.} {\bf  2015}, {\it  45},  1179-1189.

\bibitem{AS} Aspect, A.;  Dalibard, J.;  
Roger, G.,   Experimental test of Bell's Inequalities using time-varying analyzers, {\it Phys. Rev. Lett.} {\bf 1982},  {\it  49}, 1804.


\bibitem{AD1} Adenier, G., \&  Khrennikov, A.    Is the fair sampling assumption supported by EPR experiments?  
{\it J.  Phys. B: Atomic, Molecular and Optical Physics} {\bf 2007}, {\it  40}, 131-141.

\bibitem{B2} Giustina, M. et al. A significant-loophole-free test of Bell's
theorem with entangled photons. \textit{ Phys. Rev. Lett.} \textbf{2015}, \textit{115}, 250401.

\bibitem{B3} Shalm. L. K. et al. A strong loophole-free test of local
realism.  {\it Phys. Rev. Lett.} \textbf{2015}, \textit{115}, 250402.

\bibitem{B1} Hensen. B. et al, Experimental loophole-free violation of a
Bell inequality using entangled electron spins separated by 1.3 km. \textit{Nature} \textbf{2015,} \textit{526}, 682.

 \bibitem{AD2} Adenier, G.,  \&  Khrennikov, A.  Test of the no-signaling principle in the Hensen loophole-free CHSH experiment. 
{\it Fortschritte der Physik (Progress in Physics)} {\bf  2016},  {\it 65},  1600096.

 \bibitem{W} Khrennikov, A.;   Ramelow, S.;   Ursin, R.;  Wittmann, B.;  Kofler, J.;  Basieva, I. 
On the equivalence of the Clauser-Horne and Eberhard inequality based tests. {\it Physica Scripta}  {\bf 2014}, {\it  T163} 014019.


\end{thebibliography}
\end{document}